\documentclass[twocolumn,prl,longbibliography,floatfix,superscriptaddress,showkeys]{revtex4}
\usepackage{mathrsfs,bm,amsmath}
\usepackage{graphicx}
\usepackage{color}

\def\d{{\partial}}
\def\s{{\sigma}}
\def\e{{\epsilon}}

\def\0{{ {\bm 0} }}

\def\w{{\omega}}
\def\a{{\alpha}}

\allowdisplaybreaks[4]

\makeindex

\begin{document}

\title{
Edge-induced strongly correlated electronic states
in two-dimensional Hubbard model:
Enhancement of magnetic correlations and self-energy effects
}

\author{Shun Matsubara}
\author{Youichi Yamakawa}
\author{Hiroshi Kontani}
\affiliation{Department of Physics, Nagoya University, 
Nagoya 464-8602, Japan}

\date{\today}

\begin{abstract}
To understand nontrivial edge electronic states
in strongly-correlated metals such as cuprate superconductors,
we study the two-dimensional Hubbard models with open edge boundary.
The position-dependences of the spin susceptibility
and the self-energy are carefully analyzed self-consistently, by using the 
fluctuation-exchange (FLEX) approximation.
It is found that spin susceptibilities are strongly enlarged
near the (1,1) open edge when the system is near the half-filling.
The enhancement is large
even if the negative feedback from the self-energy is considered
in the FLEX approximation.
The present study predicts the emergence of nontrivial
spin-fluctuation-driven phenomena near the edge,
like the quantum criticality, edge superconductivity,
and the bond-density-wave order.

\end{abstract}



\maketitle


In strongly correlated electron systems,
many-body electronic states are drastically modified
by introducing real-space structures,
such as the defects and domain boundaries.
To predict exotic electronic properties created 
by introducing the defects in real space,
it is important to develop theoretical methods
of analyzing the strongly correlated metals without translational symmetry.
In cuprate high-$T_{\rm c}$ superconductors, for example,
single nonmagnetic impurity on Cu-site
induces the local moment with $\sim1 \mu_{\rm B}$
in both YBa$_2$Cu$_3$O$_{7-x}$ (YBCO) 
\cite{Alloul99-2}
and La$_{2-\delta}$Sr$_\delta$CuO$_4$ (LSCO)
\cite{Ishida96}.
It was revealed by the NMR study
\cite{Alloul94,Alloul00,Alloul00-2}
that both the local and the staggered spin 
susceptibilities are strongly enhanced around the impurity site.
In addition, dilute nonmagnetic impurities
cause huge residual resistivity
beyond the $s$-wave unitary scattering limit
in cuprate superconductors
\cite{Kimura}
and heavy-fermion systems
\cite{Jaccard1,Jaccard2}.
Thus, the system approaches to the magnetic 
quantum-critical point (QCP) by introducing dilute point defects.
\cite{Kontani-imp}.

In systems near the magnetic QCP without randomness,
various interesting non-Fermi liquid phenomena
are driven by spin fluctuations,
such as the $T$-linear resistivity
above the pseudo-gap temperature $T^\ast$ in cuprates
\cite{Moriya,Pines,Kontani-rev}.
It was recently revealed that 
spin fluctuations drive nontrivial ``nematic transitions'',
such as the rotational symmetry breaking at $T=T^*$
\cite{Matsuda}
and the axial charge-density-wave (CDW) formation at $T_{\rm CDW}(<T^*)$
\cite{Ghiringhelli2012sc%
,Chang:2012ib%
,Fujita:2014kg%
,Wu:2015bt%
,Hamidian:2015eo%
,Comin:2016ho},
which attract increasing attention recently.
The idea of the ``spin-fluctuation-driven CDW''
due to higher-order many-body effects (such as the vertex corrections) 
has been studied in various theoretical models.
\cite{Onari-SCVC,Tsuchiizu-Ru,Tsuchiizu-CDW,Sachdev,Metzner,Chubukov,DHLee-PNAS,Kivelson-NJP,Yamakawa-FeSe,Onari-CDW}.
For cuprates, various bond CDW order states, which are the nematic 
transitions given by the symmetry-breaking in the self-energy,
have been proposed in Refs. 
\cite{Yamakawa-CDW,Kawaguchi-CDW,Tsuchiizu-CDW}.
The ``effective hopping integrals due to self-energy'' 
have in-plane anisotropy in the bond CDW state.
Since spin fluctuations drive various fundamental phenomena,
it is significant to understand how the spin fluctuations
are modified by the real-space structures.
However, theoretical studies performed so far has been limited.

Effects of point defects in cuprates
have been studied by many theorists
 \cite{Ziegler,Poilblanc,Bulut89,Sandvik,Bulut01,Bulut00,Ohashi}.
In the random-phase-approximation (RPA),
the antiferromagnetic (AFM) spin fluctuations are enlarged
around the impurity site in the square-lattice Hubbard model,
when the impurity potential is nonlocal
 \cite{Bulut01,Bulut00,Ohashi}.
The impurity-induced enhancement of AFM fluctuations 
is obtained in the strongly correlated region
even if the impurity potential is local,
by calculating the site-dependent self-energy
based on the GV$^I$ method 
\cite{Kontani-imp}.
These results indicate that the AFM fluctuations strongly develop
near the open edge of the cluster Hubbard model,
since the edge potential is given by the impurity sites in a straight line.
However, detailed theoretical analysis 
of the ``open edge Hubbard model'' based on the 
spin fluctuation theories has not been performed yet.
(Note that the effect of the nonmagnetic impurities 
and open edges in graphene have been discussed in Refs. 
\cite{Fujita,Hirashima}.)

In the present paper,
we study the site-dependent spin susceptibility and self-energy 
in the open edge Hubbard model, by using
the RPA and the fluctuation-exchange (FLEX) approximation
\cite{Bickers}.
In both approximations,
the AFM fluctuations are found to be strongly enlarged
near the open edge, especially in the $(1,1)$ open edge model.
For this reason, both the mass-enhancement 
($Z=m^*/m=1-\d\Sigma(\e)/\d\e|_{\e=0}$)
and the quasiparticle damping 
($\gamma^*= {\rm Im}\Sigma(-i\delta)/Z$)
given by the spin-fluctuation-induced self-energy 
takes large value near the open edge.
These results indicate the emergence of exotic edge electronic states
in strongly-correlated metals,
like the quantum-critical phenomena, enhancement of superconductivity,
and spin-fluctuation-driven CDW order.

\begin{figure}[t]
\includegraphics[width=6cm]{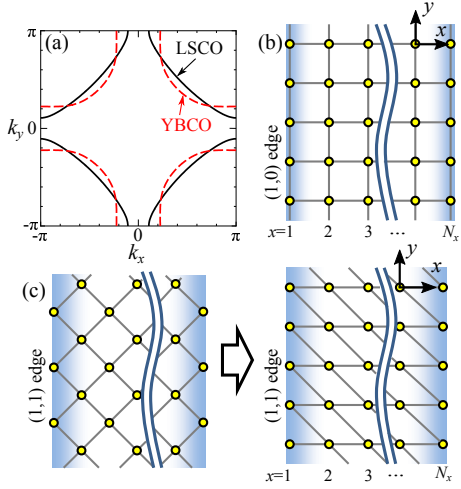}
\caption{(color online)
(a) Fermi surfaces in the YBCO and LSCO TB models at $n=0.95$.
(b)(c) Cluster models with
$(1,0)$ and $(1,1)$ open edges, respectively.
As the $(1,1)$ edge model,
we analyze the one-site unit cell structure shown in 
the right-hand-side of (c).
In (b) and (c), only the nearest-neighbor bonds connected by $t$ 
are shown by solid lines.
}
\label{fig:fig1}
\end{figure}

In this paper, we study the square-lattice cluster Hubbard model
\begin{eqnarray}
H=\sum_{i,j,\s}t_{i,j}c_{i\s}^\dagger c_{j\s}
+U\sum_{i}n_{i\uparrow}n_{i\downarrow} ,
\end{eqnarray}
where $U$ is the on-site Coulomb interaction,
and $t_{i,j}$ is the hopping integral between sites $i$ and $j$.
We set the nearest, the next nearest, the third-nearest hopping integrals
as $(t,t',t'')=(-1,1/6,-1/5)$ for YBCO tight-binding (TB) model,
and $(t,t',t'')=(-1,1/6,0)$ for LSCO TB model.
Figure \ref{fig:fig1} (a) shows the Fermi surfaces of YBCO and 
LSCO TB models for the filling $n=0.95$ without edges.
Figures \ref{fig:fig1} (b) and (c) show the cluster models with
$(1,0)$ and $(1,1)$ open edges, respectively.
In both clusters, the layer $x=1$ or $N_x$ corresponds to the edge layer.
Both models are periodic along the $y$ direction.

We analyze the site-dependent electronic states of the cluster 
Hubbard model by using the RPA and FLEX approximation.
The irreducible susceptibility is
\begin{eqnarray}
\chi^0_{x,x'}({q}_y,\w_l) &=&-T\sum_{{k}_y,n}
G_{x,x'}({q}_y+{k}_y,\w_l+\e_n)
\nonumber \\
& &\times G_{x',x}({k}_y,\e_n) ,
\label{eqn:chi0}
\end{eqnarray}
where 
$\w_l=2l\pi iT$ and $\e_n=(2n-1)\pi iT$ are the boson and
fermion Matsubara frequencies.
${\hat G}({k}_y,\e_n)=((\e_n+\mu){\hat 1}-{\hat H}_{k_y}^0
-{\hat\Sigma}({k}_y,\e_n))^{-1}$ 
is the $N_x\times N_x$ Green function.
In the RPA, the self-energy ${\hat\Sigma}({k}_y,\e_n)$ is dropped.
The spin (charge) susceptibility is given as
\begin{eqnarray}
\hat{\chi}^{s(c)}({q}_y,\w_l)=\hat{\chi}^{0}({q}_y,\w_l)
({\hat 1}-(+)U\hat{\chi}^{0}({q}_y,\w_l))^{-1} .
\label{eqn:chisc}
\end{eqnarray}
%
The Stoner factor $\a_S$ is given as the largest eigenvalue
of $U\hat{\chi}^{0}({q}_y,\w_l)$ at $\w_l=0$.
The magnetic order is realized when $\a_S\ge1$.

In the FLEX approximation, the self-energy is 
\begin{eqnarray}
\Sigma_{x,x'}({k}_y,\e_n)=T\sum_{{q}_y,l}
G_{x,x'}({k}_y+{q}_y,\e_n+\w_l)V_{x,x'}({q}_y,\w_l) ,
\label{eqn:Sigma}
\end{eqnarray}
where $\displaystyle \hat{V}({q}_y,\w_l)=U^2(\frac32\hat{\chi}^s({q}_y,\w_l)
+\frac12\hat{\chi}^c({q}_y,\w_l)-\hat{\chi}^0({q}_y,\w_l))$.
In the FLEX approximation, we solve Eqs. (\ref{eqn:chi0})-(\ref{eqn:Sigma})
self-consistently.


Hereafter, we perform the RPA and FLEX analyses for the 
cluster Hubbard models.
The $(1,0)$ edge cluster model is shown in Fig. \ref{fig:fig1} (b).
For the $(1,1)$ edge model, we analyze the one-site unit cell structure
shown in the right-hand-side of Fig. \ref{fig:fig1} (c).
In both models, we set the size of the $x$-direction as $N_x=64$,
and assume the translational symmetry along $y$-direction.
The number of $k_y$-meshes is $N_y=64$,
and the number of Matsubara frequencies is 1024 (Figs. 2-4)
or 2048 (Fig. 5).
We set the electron filling $n=0.95$, and the temperature $T=0.02$.
Here, the unit of the energy is $|t|$, which corresponds to 
$\sim 0.4$eV in cuprare superconductors without renormalization.

\begin{figure}[t]
\includegraphics[width=8cm]{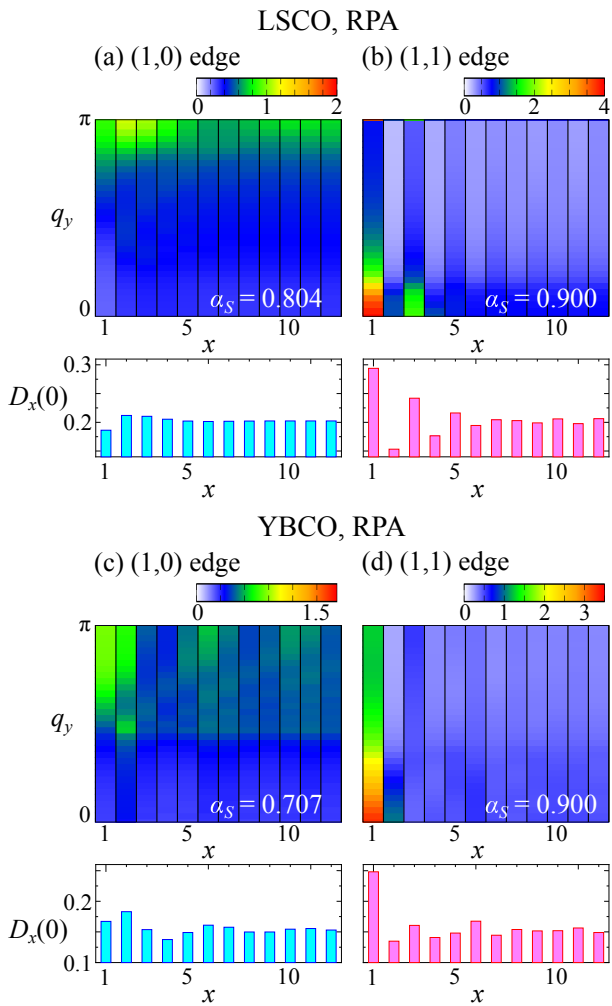}
\caption{(color online)
(a)(b) In LSCO TB model:
$\chi_{x,x}^s(q_y)$ given by the RPA 
and the LDOS at the Fermi level $D_x(0)$
for the (a) $(1,0)$ edge model
and (b) $(1,1)$ edge model, respectively.
The edge layer is $x=1$.
(c)(d) In YBCO TB model:
$\chi_{x,x}^s(q_y)$ and $D_x(0)$ for the (c) $(1,0)$ edge model
and (d) $(1,1)$ edge model, respectively.
}
\label{fig:fig2}
\end{figure}

First, we study ${\hat \chi}^s(q_y)$ at $\w_l=0$ using the RPA.
Figures \ref{fig:fig2} (a) and (b) show the static RPA susceptibilities 
$\chi_{x,x}^s(q_y)$ for the LSCO TB model at $U=1.39$.
The Stoner factor is $\a_S=0.804$ and $\a_S=0.900$
for (a) ($1,0$) edge model and (b) ($1,1$) edge model, respectively.
Since $\a_S=0.781$ in the absence of edges,
the system approaches to the magnetic QCP
by introducing the edge.
In Fig. \ref{fig:fig2} (a), $\chi_{x,x}^s(q_y)$ 
in the ($1,0$) edge model has the largest peak 
in the second layer $x=2$, at the wavevector $q_y=\pi$.
Thus, the AFM correlation increases in the second layer. 
In Fig. \ref{fig:fig2} (b), $\chi_{x,x}^s(q_y)$ 
in the ($1,1$) edge model has large peak 
in the first layer $x=1$ at $q_y=0$.
This result means that strong ferromagnetic (FM) 
fluctuations develop at the ($1,1$) open edge.
That is, the original FM correlation between the 
next-nearest-neighbor sites in the bulk 
is enlarges at the edge layer.

Figures \ref{fig:fig2} (c) and (d) show the static 
$\chi_{x,x}^s(q_y)$ for the YBCO TB model at $U=2.13$.
Here, $\a_S=0.707$ in the (c) ($1,0$) edge model, and
$\a_S=0.900$ in the (d) ($1,1$) edge model, respectively.
Since $\a_S=0.639$ in the absence of edges,
the spin fluctuations are strongly enlarged near the edge.
The obtained $(x,q_y)$-dependence of the spin susceptibility 
in the YBCO TB model
are essentially similar to those in the LSCO TB model.
(In Fig. \ref{fig:fig2} (c), $\chi_{x,x}^s(q_y)$ has 
the largest peak in the first edge layer $x=1$.)
To summarize, in the RPA,
strong magnetic fluctuations are induced near the open edge, 
insensitive to the detail of the TB model parameters.

In Figs. \ref{fig:fig2} (a)-(d),
we see that the $x$-dependence of $\chi_{x,x}^s(q_y)$
well corresponds to that Friedel oscillation of the 
local density-of-states (LDOS),
$\displaystyle D_x(\e)= \frac1{2\pi^2}\int_{-\pi}^{\pi}{d k_y} 
{\rm Im} G_{x,x}^0(k_y,\e-i\delta)$, at $\e=0$.
Thus, the edge-induced spin-fluctuation enhancement
originates from the large LDOS spot due to the 
Friedel oscillation 
\cite{Kontani-imp}.
\textcolor{black}
{
In the Supplemental Material (SM) \cite{SM},
we verify that $\chi_{x,x}^s(q_y)$ tends to become large
at which $D_x(\e)$ is large.
}

\begin{figure}[t]
\includegraphics[width=8cm]{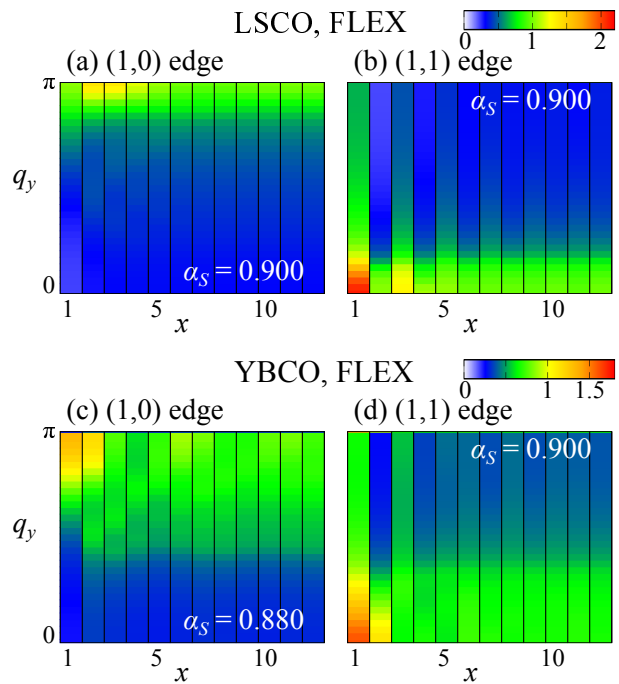}
\caption{(color online)
(a)(b) In LSCO TB model:
$\chi_{x,x}^s(q_y)$ given by the FLEX approximation
(a) $(1,1)$ edge model and (b) $(1,0)$ edge model, respectively.
(c)(d) In YBCO TB model:
$\chi_{x,x}^s(q_y)$ for the (c) $(1,0)$ edge model
and (d) $(1,1)$ edge model, respectively.
}
\label{fig:fig3}
\end{figure}

Now, we study ${\hat \chi}^s(q_y)$ using the FLEX approximation,
in order to understand the negative feedback effect 
due to the site-dependent self-energy.
Figures \ref{fig:fig3} (a) and (b) show the obtained static 
$\chi_{x,x}^s(q_y)$ in the LSCO TB model at $U=1.78$,
in the (a) ($1,0$) edge model and (b) ($1,1$) edge model.
The Stoner factor $\a_S$ is 0.900 for both (a) and (b).
Note that $\a_S=0.896$ in the absence of edges.
Figures \ref{fig:fig3} (c) and (d) show the static 
$\chi_{x,x}^s(q_y)$ in the YBCO TB model at $U=3.54$, 
in the (c) ($1,0$) edge model ($\a_S=0.880$) 
and (d) ($1,1$) edge model ($\a_S=0.900$), respectively.
Note that $\a_S=0.836$ in the absence of edges.

Therefore, the enhancement of the spin susceptibility
near the edge given by the RPA
is also verified by the FLEX approximation.
In the YBCO TB model, by introducing the $(1,1)$ edge,
$\a_S$ increases from $0.836$ ($0.641$) to $0.900$
in the FLEX approximation (RPA).
The increment of $\a_S$ becomes smaller compared to the RPA,
because of the negative feedback between $\chi^s$ and self-energy.

\begin{figure}[t]
\includegraphics[width=8cm]{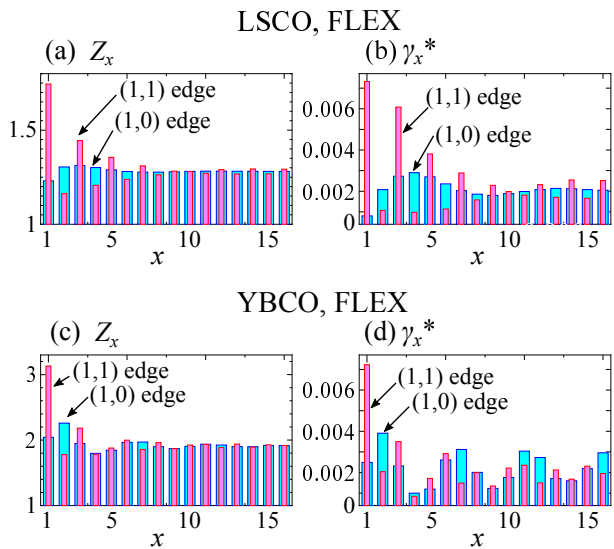}
\caption{(color online)
FLEX results in the $(1,0)$ edge and $(1,1)$ edge cluster models at $T=0.02$:
(a) Local mass-enhancement factor $Z_x$ and
(b) local quasiparticle damping rate $\gamma_x^*$ obtained in the LSCO TB model.
(c) $Z_x$ and (d) $\gamma_x^*$ obtained in the YBCO TB model.
}
\label{fig:fig4}
\end{figure}

Hereafter, we discuss the 
site-dependence of the self-energy ${\hat \Sigma}(k_y,\e-i\delta)$
given by the FLEX approximation.
First, we show the numerical results in the LSCO TB model.
Figure \ref{fig:fig4} (a) shows the 
local mass-enhancement factor
$\displaystyle Z_x =1-\frac1{2\pi}\int_{-\pi}^{\pi}dk_y
\frac{\d}{\d\e}{\rm Re}\Sigma_{x,x}(k_y,\e-i\delta)|_{\e=0}$
in LSCO at $U=1.78$.
In the $(1,0)$ edge model, $Z_x \approx 1.3$ for any $x\ (\ge1)$.
In the $(1,1)$ edge model, in contrast,
$Z_x$ increases to $1.75$ at the edge.
Figure \ref{fig:fig4} (b) shows the 
\textcolor{black}
{
local quasiparticle damping rate at Fermi energy, 
which we defined as
$\gamma_x^* \equiv \gamma_x/Z_x$, 
where $\displaystyle \gamma_x=\frac1{2\pi}\int_{-\pi}^{\pi}dk_y
{\rm Im}\Sigma_{x,x}(k_y,0-i\delta)$.
}
In the $(1,0)$ edge model, the site-dependence of $\gamma_x^*$ is moderate.
In contrast, $\gamma_x^*$ at the edge ($x=1$) takes large value 
in the $(1,1)$ edge model, due to the strong spin fluctuations 
near the edge.

Next, we show the numerical results in the YBCO TB model.
Figure \ref{fig:fig4} (c) shows the obtained 
$Z_x$ in YBCO at $U=3.54$.
In the $(1,1)$ edge model, $Z_x$ increases from $2$
in the bulk to $3.2$ at the edge.
Figure \ref{fig:fig4} (d) shows the obtained $\gamma_x^*$.
In both $(1,0)$ and $(1,1)$ edge models
$\gamma_x^*$ increases near the edge layer.
In the $(1,1)$ edge model, $\gamma_x^*$ drastically 
increases to $0.0072$ at the edge layer.

\begin{figure}[t]
\includegraphics[width=8cm]{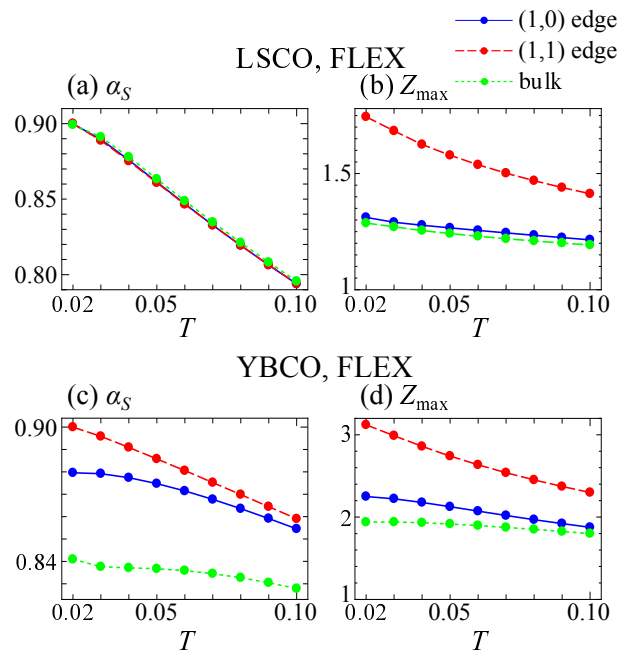}
\caption{(color online)
$T$-dependences of the 
FLEX results in the $(1,0)$ edge and $(1,1)$ edge cluster models,
and in bulk model (without edges).
(a) Stoner factor $\a_S$ and (b) maximum local 
mass-enhancement factor $Z_{\rm max}$ obtained in the LSCO TB model.
(c) $\a_S$ and (d) $Z_{\rm max}$ obtained in the YBCO TB model.
}
\label{fig:fig5}
\end{figure}

Therefore, by introducing the open edge  
in metals with moderate AFM fluctuations ($\a_S\sim0.9$),
strong AFM or FM fluctuations are induced near the open edge.
The induced strong spin fluctuations 
give rise to huge quasiparticle damping rate
and mass-enhancement near the open edge.
The present study indicates that various extreme 
quantum critical phenomena are expected to emerge 
near the open edge.


Finally, we examine the $T$-dependences of the 
electronic states in detail based on the FLEX approximation.
Figures \ref{fig:fig5} (a)-(d) show the Stoner factor $\a_S$
and the largest local mass-enhancement factor 
$\displaystyle Z_{\rm max}={\rm max}_x \{Z_x\}$.
The results in LSCO model are shown in Figs. \ref{fig:fig5} (a) and (b):
Although $\a_S$ is insensitive to open edges,
$Z_{\rm max}$ strongly increases as $T$ decreases
near ($1,1$) edge by reflecting the large $\chi^s$
at the edge layer shown in Fig. \ref{fig:fig3} (b).
In YBCO model, as shown in Figs. \ref{fig:fig5} (c) and (d),
both $\a_S$ and $Z_{\rm max}$ strongly increase 
as $T$ decreases in the presence of ($1,1$) edge.
These results mean the emergence of interesting 
edge-induced quantum critical phenomena.


\textcolor{black}
{
In the SM \cite{SM}
we present the numerical results for $n=0.90\sim1.10$,
and find that prominent edge-induced quantum criticality 
appears when the edge LDOS is large. 
This result is an useful guideline to realize
the quantum criticality driven by real space structures.
The large damping may be observed experimentally, 
as the pseudo-gap formation in the LDOS in (1,1) open edge.
}

In summary,
we studied the site-dependent spin susceptibility and self-energy 
in the open edge Hubbard model.
The magnetic fluctuations are found to be strongly enlarged
near the open edge, especially for the $(1,1)$ edge case.
In the FLEX, both the local mass-enhancement factor $Z_x$
and the local quasiparticle damping $\gamma_x^*$
given by the spin-fluctuation-induced self-energy 
become huge near the open edge.
Thus, interesting edge-induced quantum critical phenomena
are predicted by the present study.
We note that the impurity-driven enhancements of $\chi^s$ and $\gamma^*$
are underestimated in the FLEX, since the negative feedback
between $\chi^s$ and $\Sigma$ is overestimated \cite{Kontani-imp}.
To overcome this problem,
the GV$^I$ method will be useful, since this method can successfully 
explain the impurity-induced magnetic quantum-critical phenomena 
in cuprate superconductors
\cite{Kontani-imp}.
This is one of our important future problems.
Another important future problem is to study the spin-fluctuation-driven 
nematicity discussed in Refs. 
\cite{Onari-SCVC,Onari-CDW,Kawaguchi-CDW,Tsuchiizu-CDW}.
The present study indicates the emergence of the 
impurity- or edge-induced nematic orders, 
which are actually observed in several Fe-based superconductors 
\cite{Davis,Li}.

\acknowledgements
This work was supported by Grant-in-Aid for Scientific Research from 
the Ministry of Education, Culture, Sports, Science, and Technology, Japan.



\clearpage

\makeatletter

\renewcommand{\thefigure}{S\arabic{figure}}
\renewcommand{\theequation}{S\arabic{equation}}
\makeatother
\setcounter{figure}{0}
\setcounter{equation}{0}
\setcounter{page}{1}
\setcounter{section}{1}

\begin{widetext}
\begin{center}
{\Large [Supplementary Material]}
\end{center} 

\begin{center}
{\large
\textbf{
Edge-induced strongly correlated electronic states
in two-dimensional Hubbard model:
Enhancement of magnetic correlations and self-energy effects
}}
\end{center} 
\begin{center}
Shun Matsubara,
Youichi Yamakawa, and
Hiroshi Kontani
\end{center} 


\end{widetext}

In the main text, we present only the numerical results 
of the electron filling $n=0.95$, which corresponds to 
under-doped region of hole-doping compounds.
Interesting edge-induced quantum critical phenomena are realized 
in both LSCO TB and YBCO TB cluster Hubbard models.
In this supplemental material (SM),
we present the numerical results for $n=0.90\sim1.10$
in order to understand the origin of the edge-induced quantum criticality.
We find the realization condition of the prominent
edge-induced quantum criticality.

\subsection{A: filling dependence in LSCO TB model}

First, we explain the numerical results for the LSCO TB model.
We set $T=0.02$ and $U=1.78$ in unit eV,
for $n=0.90$, 0.95, 1.05 and 1.10.
Figure \ref{fig:figS1} presents the (a) $\chi_{x,x}^s(q_y)$
and (b) quasiparticle damping $\gamma_x^*$, 
mass-enhancement factor $Z_x$, 
and the bare local density of states (LDOS) at the Fermi level $D_x(0)$.
The obtained spin Stoner factors are shown in 
Fig. \ref{fig:figS1} (a).
The oscillation in $D_x(0)$ is understood as 
the Friedel oscillation caused by the open edge.


In the $(1,1)$ edge model, 
the edge-induced quantum criticality is prominent
in both hole-doped case ($n<1$) and electron-doped case ($n>1$).
$\chi_{x,x}^s(q_y)$ is strongly enlarged at $x=1$ and $q_y=0$.
By reflecting this fact, both $\gamma_x^*$ and $Z_x$
are strongly enlarged in both hole- and electron-doped cases.
In the $(1,0)$ edge model,
the edge-induced quantum criticality is moderate.
In electron-doped case, $\chi_{x,x}^s(q_y)$ takes the maximum 
at $x=1$ and $q_y=\pi$.
In hole-doped case, in contrast, $\chi_{x,x}^s(q_y=\pi)$
is moderately enlarged at $x\ge2$.
Both $\gamma_x^*$ and $Z_x$ show similar $x$-dependences
to $D_x(0)$.

The obtained nontrivial $n$-dependences for both
$(1,1)$ and $(1,0)$ edge models are well understood 
in terms of the LDOS without interaction shown in 
Fig. \ref{fig:figS1} (b).
In the $(1,0)$ edge model,
the LDOS at $x=1$ is strongly suppressed in hole-doped case.
Due to this fact, the edge electronic states deviate from the 
quantum criticality.
In electron-doped case, the LDOS at $x=1$ is larger than the balk DOS,
so the edge effect becomes moderate.
In the $(1,1)$ edge model,
the $x$-dependence of the LDOS is essentially $n$-independent.
For this reason, the edge electronic states approach to the 
quantum criticality in both hole-doped and electron-doped cases.

\subsection{B: filling dependence in YBCO TB model}

Next, we explain the numerical results for the YBCO TB model
for $n=0.90$, 0.95, 1.05 and 1.10.
The obtained $\chi_{x,x}^s(q_y)$, $\gamma_x^*$, $Z_x$, and $D_x(0)$
are shown in Fig. \ref{fig:figS2}.
The YBCO TB model with $n>1$ corresponds to 
the electron-doped cuprate superconductors, NCCO and PCCO.
The obtained spin Stoner factors are shown in 
Fig. \ref{fig:figS2} (a).

In both $(1,0)$ and $(1,1)$ edge models,
the obtained $n$-dependences are qualitatively similar to 
those obtained in the LSCO TB model.
In the $(1,1)$ edge model, 
$\chi_{x,x}^s(q_y=0)$,  are 
strongly enlarged at $x\approx1$.
Thus, the edge electronic states approach to the quantum criticality.
This result originates from the large LDOS on the $(1,1)$ edge
in YBCO model, shown in Fig. \ref{fig:figS2} (b).
In the $(1,0)$ edge model, 
the edge-induced quantum criticality is moderate.
For both $n>1$ and $n<1$ cases,
$\chi_{x,x}^s(q_y=\pi)$ takes the maximum at $x=1$.
In hole-doped case, in contrast $\chi_{x,x}^s(q_y=\pi)$
is moderately enlarged at $x\ge2$.
Both $\gamma_x^*$ and $Z_x$ show similar $x$-dependences.

To summarize, prominent edge-induced quantum criticality 
is realized when the edge LDOS is large. 
This result is an useful principle to control the 
quantum criticality driven by real space structure,
since it is easy to calculate the LDOS in non-interacting systems.
The YBCO TB model with $n>1$ corresponds to NCCO and PCCO.
In YBCO TB model,
very large quasiparticle damping rate $\gamma_x^*$
is obtained in the (1,1) open edge.
This result may lead to the pseudo-gap formation in
the LDOS in the (1,1) open edge in YBCO, NCCO, and PCCO
cuprate superconductors.

\begin{widetext}
\begin{center}
\begin{figure}[t]
\includegraphics[width=14cm]{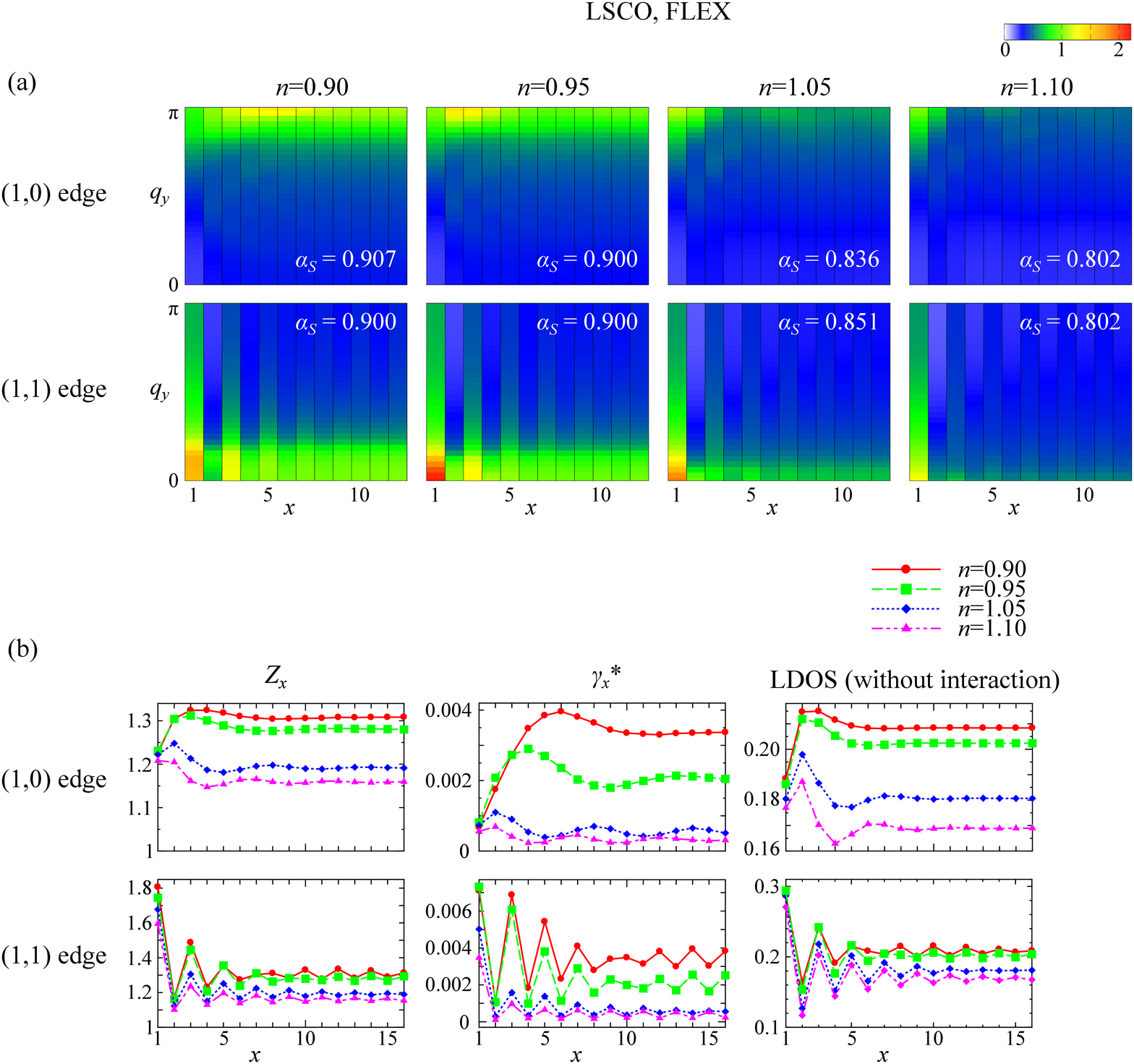}
\caption{(color online)
(a) Obtained $\chi_{x,x}^s(q_y)$ in LSCO TB model with
$(1,1)$ open edge and $(1,0)$ open edge, respectively.
The results for $n=0.90$, 0.95, 1.05, and 1.10 are shown.
(b) Obtained filling-dependences of the 
mass-enhancement factor $Z_x$ and
quasiparticle damping $\gamma_x^*$, 
and LDOS at the Fermi level $D_x(0)$
in the LSCO TB model $(n=0.90\sim1.10)$.
}
\label{fig:figS1}
\end{figure}

\begin{figure}[t]
\includegraphics[width=14cm]{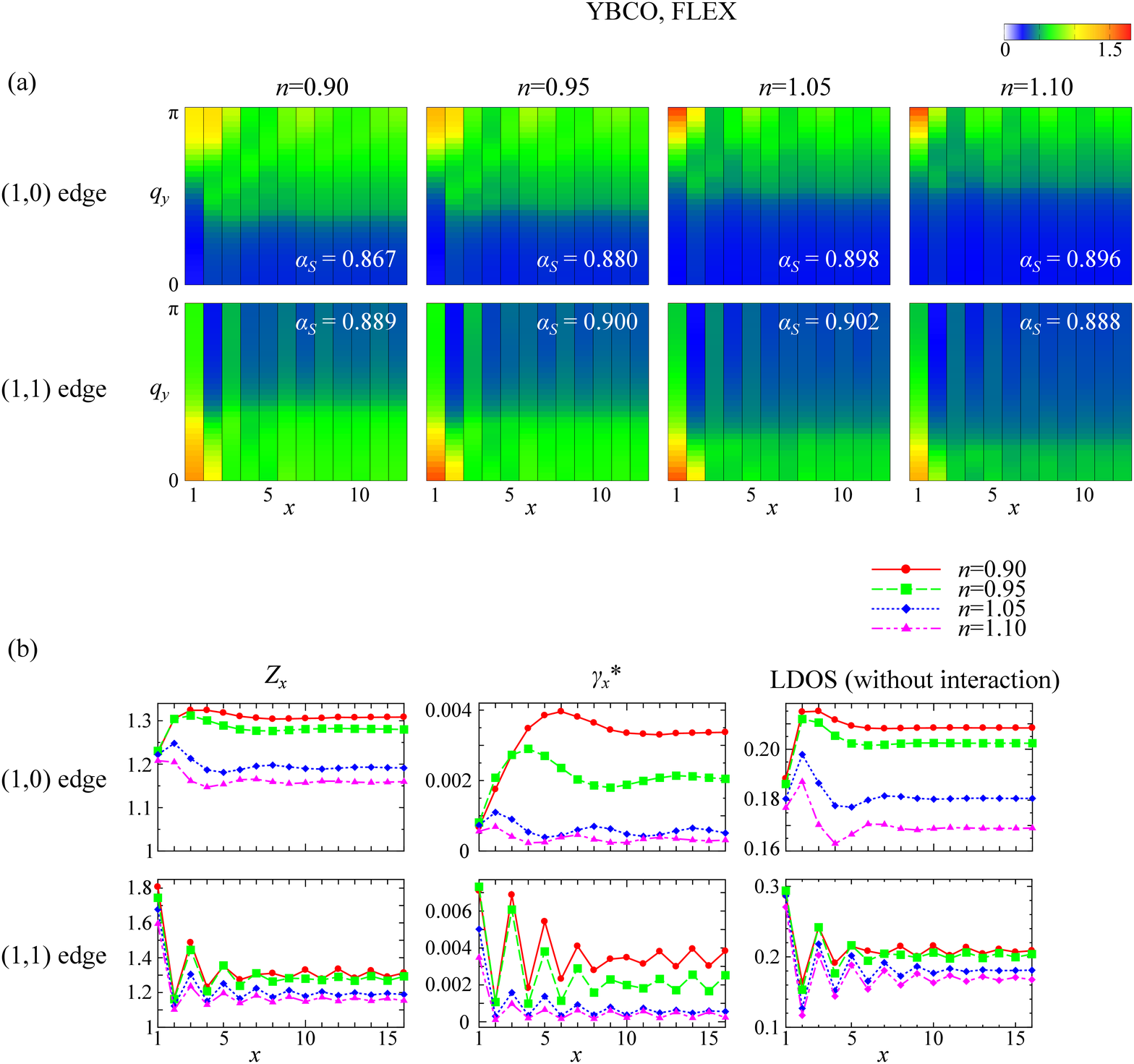}
\caption{(color online)
(a) Obtained $\chi_{x,x}^s(q_y)$ in YBCO TB model with
$(1,1)$ open edge and $(1,0)$ open edge, respectively.
The results for $n=0.90$, 0.95, 1.05, and 1.10 are shown.
(b) Obtained filling-dependences of the 
mass-enhancement factor $Z_x$ and
quasiparticle damping $\gamma_x^*$, 
and LDOS at the Fermi level $D_x(0)$
in the YBCO TB model $(n=0.90\sim1.10)$.
}
\label{fig:figS2}
\end{figure}
\end{center}
\end{widetext}

\end{document}